\documentclass{PoS}
\usepackage{booktabs}

\title{Run-Wise Simulations for Imaging Atmospheric Cherenkov Telescope Arrays}

\ShortTitle{Run-Wise Simulations for Imaging Atmospheric Cherenkov Telescope Arrays}

\author{\speaker{M. Holler}$^{a}$, J. Chevalier$^{b}$, J.-P. Lenain$^{c}$, M. de Naurois$^{d}$, D. Sanchez$^{b}$\\
        ${}^{a}$ Universit\"at Innsbruck\\
        ${}^{b}$ LAPP, Universit\'e Savoie Mont-Blanc\\
        ${}^{c}$ LPNHE, Paris\\
        ${}^{d}$ Laboratoire Leprince-Ringuet - \'Ecole Polytechnique\\
        E-mail: \email{markus.holler@uibk.ac.at}}

\abstract{We present a new paradigm for the simulation of arrays of Imaging Atmospheric Cherenkov Telescopes (IACTs) which overcomes limitations of current approaches. Up to now, all major IACT experiments rely on the same Monte-Carlo simulation strategy, using predefined observation and instrument settings. Simulations with varying parameters are generated to provide better estimates of the Instrument Response Functions (IRFs) of different observations. However, a large fraction of the simulation configuration remains preserved, leading to complete negligence of all related influences. Additionally, the simulation scheme relies on interpolations between different array configurations, which are never fully reproducing the actual configuration for a given observation. Interpolations are usually performed on zenith angles, off-axis angles, array multiplicity, and the optical response of the instrument. With the advent of hybrid systems consisting of a large number of IACTs with different sizes, types, and camera configurations, the complexity of the interpolation and the size of the phase space becomes increasingly prohibitive. Going beyond the existing approaches, we introduce a new simulation and analysis concept which takes into account the actual observation conditions as well as individual telescope configurations of each observation run of a given data set. These run-wise simulations (RWS) thus exhibit considerably reduced systematic uncertainties compared to the existing approach, and are also more computationally efficient and simple. The RWS framework has been implemented in the H.E.S.S. software and tested, and is already being exploited in science analysis.}

\FullConference{35th International Cosmic Ray Conference - ICRC2017\\
		12-20 July, 2017\\
		Bexco, Busan, Korea}

\begin{document}

\section{Introduction}

In the past years, the field of ground-based gamma-ray astronomy using Imaging Atmospheric Cherenkov Telescopes (IACTs) has undergone several improvements. Instrument upgrades both enlarge the accessible energy range and improve the overall data quality (\cite{2016_HessUpgrade}, \cite{2011_VeritasUpgrade}, \cite{2016_MagicUpgrade}). As for the data analysis, advanced photon reconstruction and analysis techniques led to improved event classification and reconstruction precision (\cite{2009_Model}, \cite{2014_ImPACT}, \cite{2009_TMVA}, \cite{2015_VeritasModel}).

Opposed to these advancements, the overall Monte-Carlo (MC) simulation and analysis strategy has up to now remained unchanged. The principle of this approach is identical for all major IACT experiments (see, e.g., \cite{2006_HessCrab}, \cite{2016_MAGICCrab}
). In general, MC simulations are carried out for predefined and fixed observation and instrument parameters. For those parameters which are assumed to have a major influence on the observation, simulations with varying parameter values are generated to provide better estimates of the Instrument Response Functions (IRFs). The remaining part of the simulated configuration is left unchanged. To obtain the response for a given observation, the IRFs for the different configurations are interpolated. The interpolation helps to achieve a better estimate, but is in most cases just a simplification and thus inevitably introduces systematic errors. Additionally, upgraded and future IACT arrays become increasingly complex because of different IACT sizes, types, and camera configurations. Thus the largely increased phase space makes it more and more difficult to run the simulations computationally efficient whilst keeping up with the nominal performance gain.

Here we introduce a new MC simulation and analysis concept as an alternative to the existing approaches. It takes into account the actual observation conditions as well as individual telescope configurations of each observation of a given data set.


\section{Principles and Technical Implementation}

The main motivation for the new approach was to allow simulations that are as close as possible at the observational reality. Most observation conditions can in general be assumed to be constant for the time scales of a unit of continuous data taking. 
As for H.E.S.S., such an observation unit lasts up to $28\,\mathrm{min}$ and is called \textit{observation run}, or just \textit{run} \cite{2006_HessCrab}. For the method presented here, dedicated simulation sets are generated on a \textit{run}-by-\textit{run} basis, therefore it is henceforth called the Run-Wise-Simulation (RWS) concept.

\subsection{Technical Framework}
\label{sec_tech}

The RWS approach has been fully implemented in one of the two analysis chains in operation in H.E.S.S., named \textit{parisanalysis}. All information that is relevant for the simulation of a specific \textit{run} is stored in a MySQL database (DB) 
and read out on demand. Hence no read-in of \textit{run}-specific files is necessary. The shower simulations are carried out with the KASCADE software \cite{1994_Kaskade}, which has been improved and rewritten in C$++$ to allow more flexible usage and easy integration in the simulation chain (see \cite{2009_Model}). It has been further enhanced for the present work to be able to simulate the evolution of the tracking position of the telescopes within a \textit{run}. The KASCADE output is directly passed to the internal IACT simulation software. Opposed to the classical simulation approach, where generally the shower and IACT simulation are decoupled, no direct output of the shower simulation is saved in case of the implemented RWS approach, leading to a drastic reduction of needed disk space at the expense of increased computation time. To allow simulating larger statistics, simulations can be split into several slices for each \textit{run}. The simulated raw-data files contain all information that is needed for the event reconstruction and higher-level analysis frameworks. A MySQL DB table is used to keep track of the simulation production, including details of the actual simulation parameters of each \textit{run}.

\subsection{Simulation Parameter Settings}

Observations of IACT arrays generally cover a large range of observation conditions, implying substantial variations of the required simulation parameter space. For an efficient simulation with yet sufficient parameter coverage for each \textit{run}, an automatized parameter calculation approach is indispensable. The relevant inputs for the calculation presented here are:
\begin{itemize}
\item Array configuration,
\item Zenith angle range $\left[ \theta_{\mathrm{min}}, \theta_{\mathrm{max}} \right]$ covered within the \textit{run},
\item Relative optical efficiencies $\epsilon$ of the participating IACTs,
\item Desired statistics level, defined by the differential flux at $E = 1\,\mathrm{TeV}$,
\item Spectral index $\Gamma$ of the simulation,
\item Number of computing jobs (slices) per \textit{run}.
\end{itemize}
The first three are \textit{run}-dependent and automatically read out from the database, whereas the three latter are set by the user.

MC simulations were carried out to determine up to which impact distances (with respect to the center of the array) and down to which primary photon energies electromagnetic showers can trigger the H.E.S.S. telescopes. Based on these, the maximum simulated impact distance is set to
\begin{equation}
R_{\mathrm{sim}}\left( \theta_{\mathrm{max}} \right) = R_{0} \cdot \frac{1}{\cos \theta_{\mathrm{max}}},
\label{Eq_Impact}
\end{equation}
where $R_{0} = 480\,\mathrm{m}$. Both the value and the functional relation are notably independent of other observation parameters, which was verified with the underlying simulations. A potential future improvement would be to make it dependent on the energy, since for photons with lower energy less simulation area is needed.
The minimum simulated primary photon energy of a \textit{run} is
\begin{equation}
E_{\mathrm{min}} = \frac{E_{0}}{\epsilon_{\mathrm{max}}}\cdot \frac{1}{\cos^3 \theta_{\mathrm{min}}},
\label{Eq_Energy}
\end{equation}
with $E_{0} = 5\,\mathrm{GeV}$ for \textit{runs} where the large H.E.S.S. telescope CT5 (\cite{2014_CT5}) took part in the observation and $E_{0} = 30\,\mathrm{GeV}$ for all other \textit{runs}. $\epsilon_{\mathrm{max}}$ corresponds to the maximum of the relative optical efficiencies of the observing IACTs, or directly to the one of CT5 if participating. Eq.~\ref{Eq_Energy} takes into account both the fact that at larger zenith angles the photon density on ground is lower due to the enlarged Cherenkov cone (see Eq.~\ref{Eq_Impact}) as well as the increased attenuation due to the larger optical depth. The maximum energy $E_{\mathrm{max}}$ of the simulated primary photons is fixed to $100\,\mathrm{TeV}$ for all observations.

As stated above, the user sets the spectral index of the simulation and the differential flux at $E = 1\,\mathrm{TeV}$. The latter is set in units of the Crab, where the reference spectrum of the Crab Nebula is taken from \cite{2015_MAGICCrab}. Together with Eq.~\ref{Eq_Impact}, \ref{Eq_Energy}, and $E_{\mathrm{max}}$, the simulation phase space for a given \textit{run} is set, allowing to derive the number of simulated events. 
\begin{figure}
\center
\includegraphics[width = 0.6\textwidth]{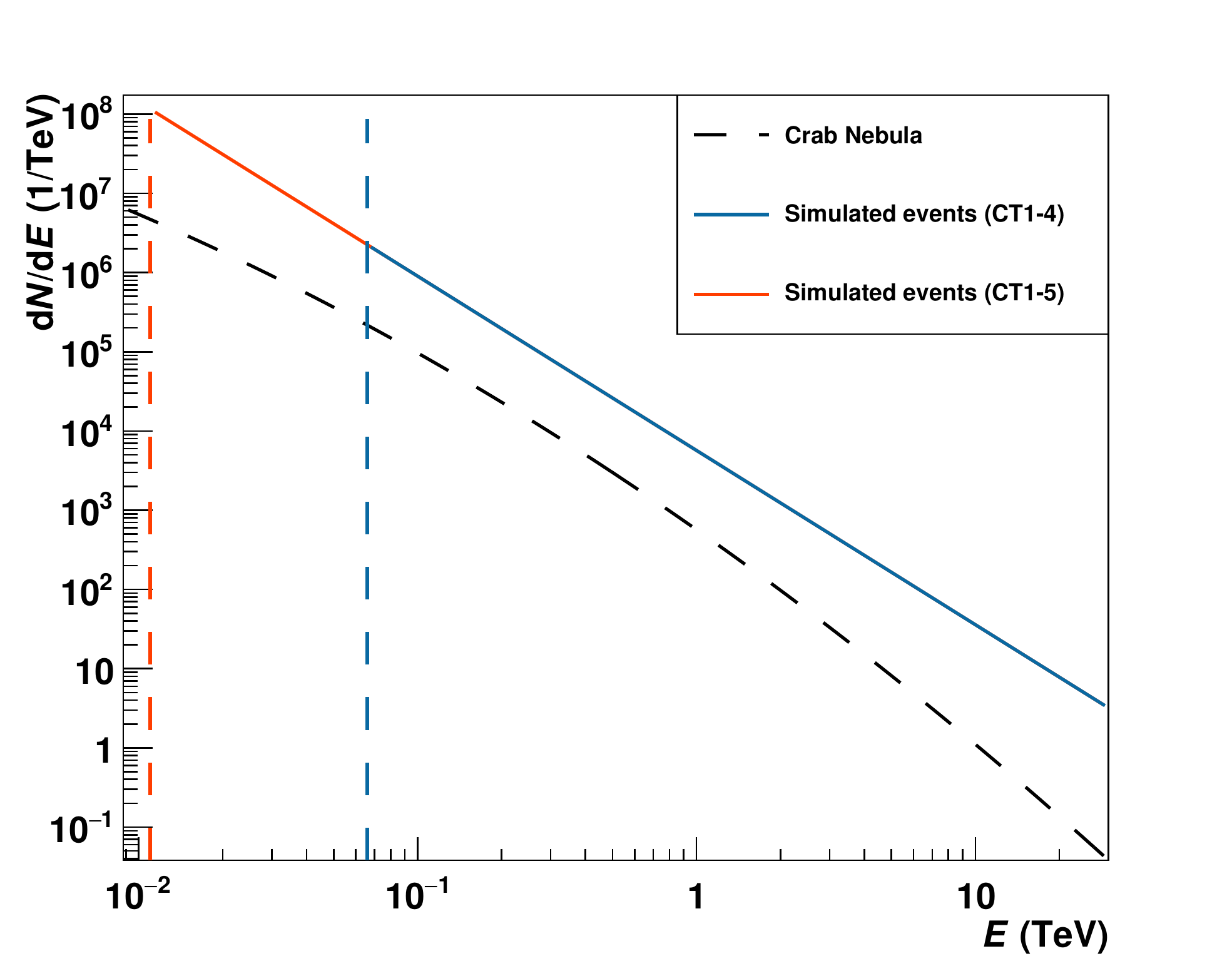}
\caption{Number of simulated events for a \textit{run} with $\theta = 30^{\circ}$, $\epsilon_{\mathrm{max}} = 70\%$, a differential spectral index $\Gamma = 2.2$, and a differential flux of $10\,$Crab at $E = 1\,$TeV. The values of $E_{\mathrm{min}}$ are illustrated with dashed red and blue lines for a run with and without CT5, respectively.}
\label{SimEvts}
\end{figure}
The event statistics per energy for the simulation of a given \textit{run} are illustrated in Fig.~\ref{SimEvts}. The parameters are $\theta = 30^{\circ}$, $\epsilon_{\mathrm{max}} = 70\%$, $\Gamma = 2.2$, and a differential flux of $10\,$Crab at $E = 1\,$TeV. The total number of simulated events is $N_{\mathrm{CT}1-4} \approx 1.2\times 10^{5}$ and $N_{\mathrm{CT}5} \approx 1.1\times 10^{6}$.

\subsection{Observation-Based Simulation Configuration}

The RWS framework exploits a large variety of information of each observation \textit{run} to enhance the description of the simulation, as is going to be laid out in the following. As previously mentioned in Section~\ref{sec_tech}, this information is stored in the DB and read out for the simulation.

\begin{table}
\caption{List of parameters that are used for the simulation of a \textit{run}.}
\label{tab_simpar}
\begin{tabular}{lcc} \toprule
Quantity & Type & Comment \\
			   & (per pixel, telescope, or for array) & \\  \midrule
Active IACTs & - & - \\
Telescope Tracking & array & see text \\
Source Position & array & see text \\
Optical Efficiency $\epsilon$ & telescope  & - \\
Transparency Coefficient & array & see text \\
Camera Focus & telescope & only relevant for CT5 \\ 
Trigger Settings & telescope & - \\ 
Live-time fraction & telescope & used for camera dead-time \\
Broken Pixels & pixel & for High and Low Gain \\
PMT Gain & pixel & - \\
Hi-Lo Ratio & pixel & - \\
Flatfield Coefficient & pixel & - \\
Night-Sky Background & pixel & see text \\
 \bottomrule
\end{tabular}
\end{table}
An overview of the \textit{run} information that is used for the simulation is given in Table~\ref{tab_simpar}. The correct use of all input parameters has been thoroughly and successfully checked. It is clear that the quality of the RWS strongly depends on a working calibration framework, which serves as the source of most of the parameters. 

\begin{figure}
\center
\includegraphics[width = \textwidth]{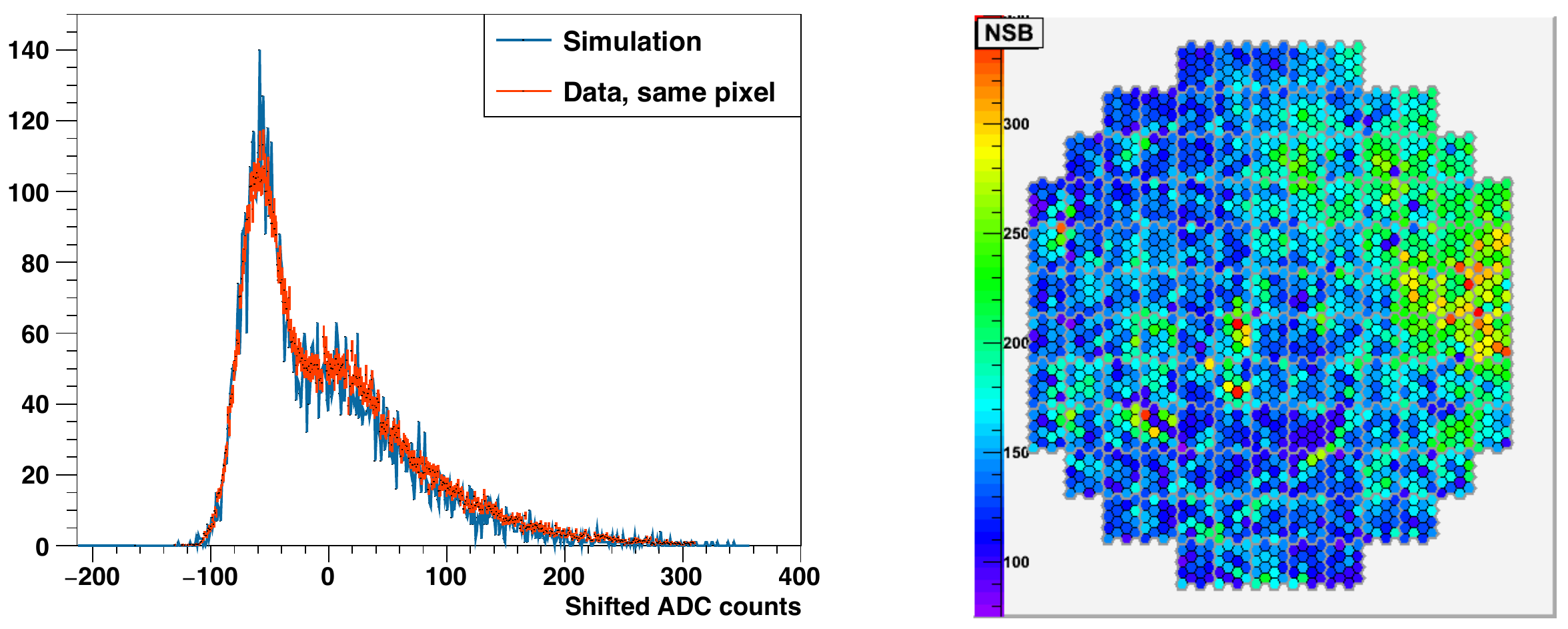}
\caption{\textit{Left:} Pedestal shapes of a specific pixel with a measured NSB rate of $47\,$MHz as a function of the charge (centered around 0). The blue curve corresponds to the measured charge histogram of the actual observation, and the red one shows the result of the simulation. \textit{Right:} Measured NSB map of CT5 for an observation of the Galactic Centre. The values that are denoted on the color bar are given in MHz/pixel.}
\label{nsb}
\end{figure}
As an example, the overall agreement of the measured and simulated pedestal shape of a given pixel for a specific \textit{run} is shown in the \textit{left} panel of Figure~\ref{nsb}. This agreement is the result of a fully working calibration and simulation chain: For each pixel, the rate of the night-sky background (NSB) is measured and stored in the DB. For the RWS, this value is again read out from the DB, and NSB photons with that rate are simulated. Whereas for \textit{standard} MC simulations the NSB rate is generally set constant throughout the FoV, even very inhomogeneous NSB fields (such as the one around the Galactic Centre in the \textit{right} panel of Figure~\ref{nsb}) are correctly simulated in our approach. 

To prepare the shower simulation, the start and end times of the respective \textit{run} are read out from the DB. The $N$ events to be simulated are equally distributed over the time window to simulate a constant source, and a unique time stamp is assigned to each event. In case the simulation of a \textit{run} is split into more than one slice, the time window of the \textit{run} is splitted accordingly. The input coordinates to be used for the simulation are given in J2000 sky coordinates. Two sets of coordinates are set and fixed throughout the \textit{run}: The pointing position and the source position. The source position can either be set manually or read out automatically for \textit{runs} where a specific source was observed. If diffuse simulations shall be generated, a diffuse cone angle around the source position is set additionally, and the directions of the simulated particles are distributed uniformly within this circle. Arbitrary source shapes can be generated by cutting in at analysis level. 
At the beginning of the simulation of each particle, the pointing direction of the telescopes is updated to simulate the trajectory of the source on the sky. The directional origin of the particle is converted accordingly. This approach corresponds to a complete, realistic simulation of the movement of a constant source in the night sky which is being tracked by an array of IACTs. 
To mimic pointing uncertainties, the position of each IACT is offset constantly throughout the \textit{run} according to a 2D Gaussian with $\sigma = 15''$. 

To account for variations of the atmospheric quality, the Cherenkov transparency coefficient \cite{2014_TC} is read out from the DB. Instead of applying it to the photon attenuation during the shower simulation process directly, it is multiplied to the simulated relative optical efficiency of each participating IACT.


\subsection{High-level Analysis}

The high-level analysis analysis framework has been adapted to work with RWS. This implies several analysis modules, which is explained in the following. 

\subsubsection{RWS MC Analysis}
\label{mc_ana}

A data set of RWS can be analysed in the same manner as actual data. A list of previously simulated \textit{runs} is passed to the analysis framework, and a test position is chosen in sky coordinates. This usually corresponds to the catalogue position of the source. As a proper reconstructed sky coordinate is assigned to each event, the generation of advanced, sky-coordinate-based diagnostic plots and even simulated acceptance maps is technically already possible.

For the MC analysis, the simulations can be re-weighed to a user-defined spectral shape. Up to now, power-law, log-parabola, and exponential cut-off power-law shapes are implemented. The user chooses between using the full available event statistics or alternatively a differential flux to re-weigh to. Both event weighing or event throwing has been implemented.

\subsubsection{Morphology Fitting}
\label{morphology}

Providing a more realistic simulation approach, RWS are very well suited for the extraction of the instrument point spread function (PSF), thus allowing more precise morphological studies. 

With the classical simulation approach, lookup tables are normally generated to store the PSF for different varying observation configurations. When using RWS, this is not anymore necessary as the PSF of a given source can be directly simulated (see Section~\ref{mc_ana}). The current approach is such as to run a MC analysis with a fine binning of the $\vartheta^2$ histogram, where the typical bin width is set to $2\cdot 10^{-4}\,\mathrm{deg}^2$. This resulting MC $\vartheta^2$ histogram is then used as the reference point-source PSF for the morphology fitting. The functionality has been fully implemented and successfully tested.

\begin{figure}
\center
\includegraphics[width = \textwidth]{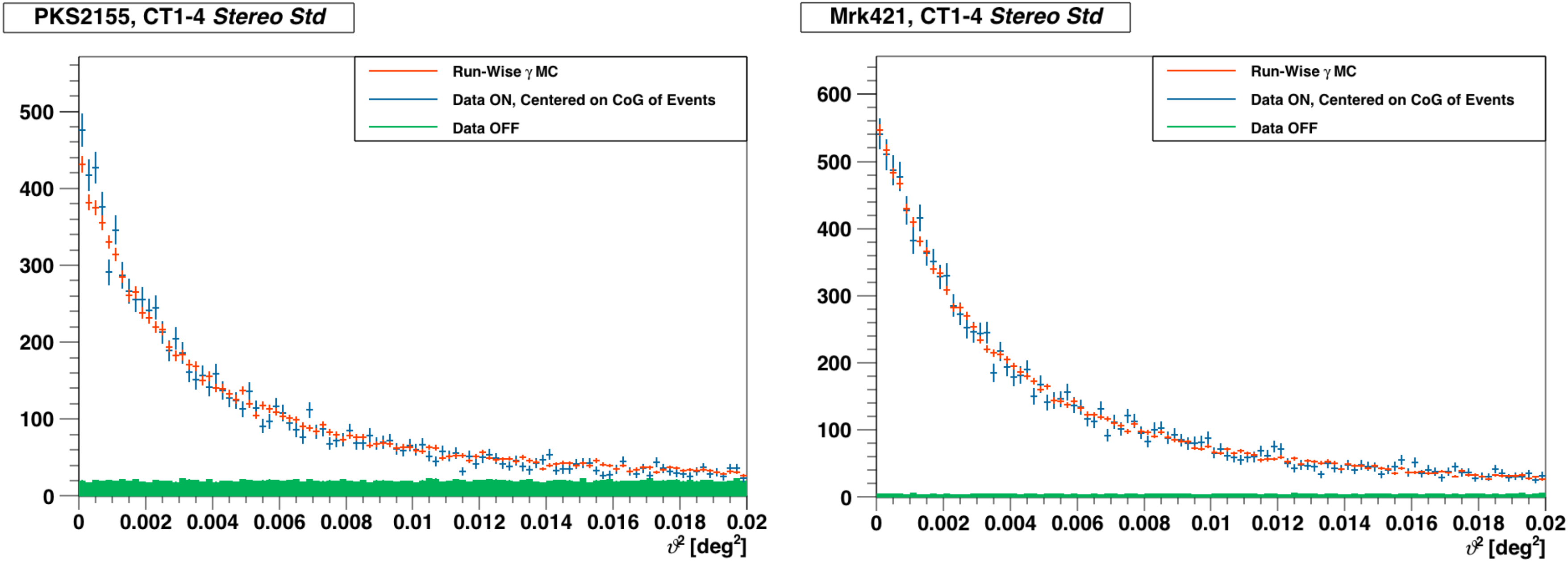}
\caption{MC-Data comparisons of the squared angular distance ($\vartheta^2$) to the source position for PKS 2155$-$304 (\textit{left}) and Mrk 421 (\textit{right}). On both plots, the blue and green histograms denote measured ON and OFF events (the latter scaled down to the size of the ON region). The red histograms correspond to the respective PSFs, obtained using matching RWS which are re-weighed to the spectral shape of the sources.}
\label{theta2}
\end{figure}
Simulated and measured $\vartheta^2$ histograms of PKS 2155$-$304 and Markarian 421 
are shown on the \textit{left} and \textit{right} panel of Fig.~\ref{theta2}, respectively. For both sources, an excellent MC-Data agreement is obtained. Without having to assume systematic uncertainties, the morphology fit yields no extension for both. The $[1\sigma,3\sigma]$ upper limits on a potential Gaussian extension width are $[13.7'',23.0'']$ for PKS 2155$-$304 and $[23.4'',33.5'']$ for Markarian 421, respectively. 

\subsubsection{Spectral Fitting}

Spectral fit results obtained with IRFs from RWS are expected to be more precise because of smaller systematic uncertainties. 

The IRF generation with RWS MCs has been adapted to the software framework. IRFs are filled and read out on a per-run basis, hence again no interpolation of lookup tables is necessary.

\begin{figure}
\center
\includegraphics[width = \textwidth]{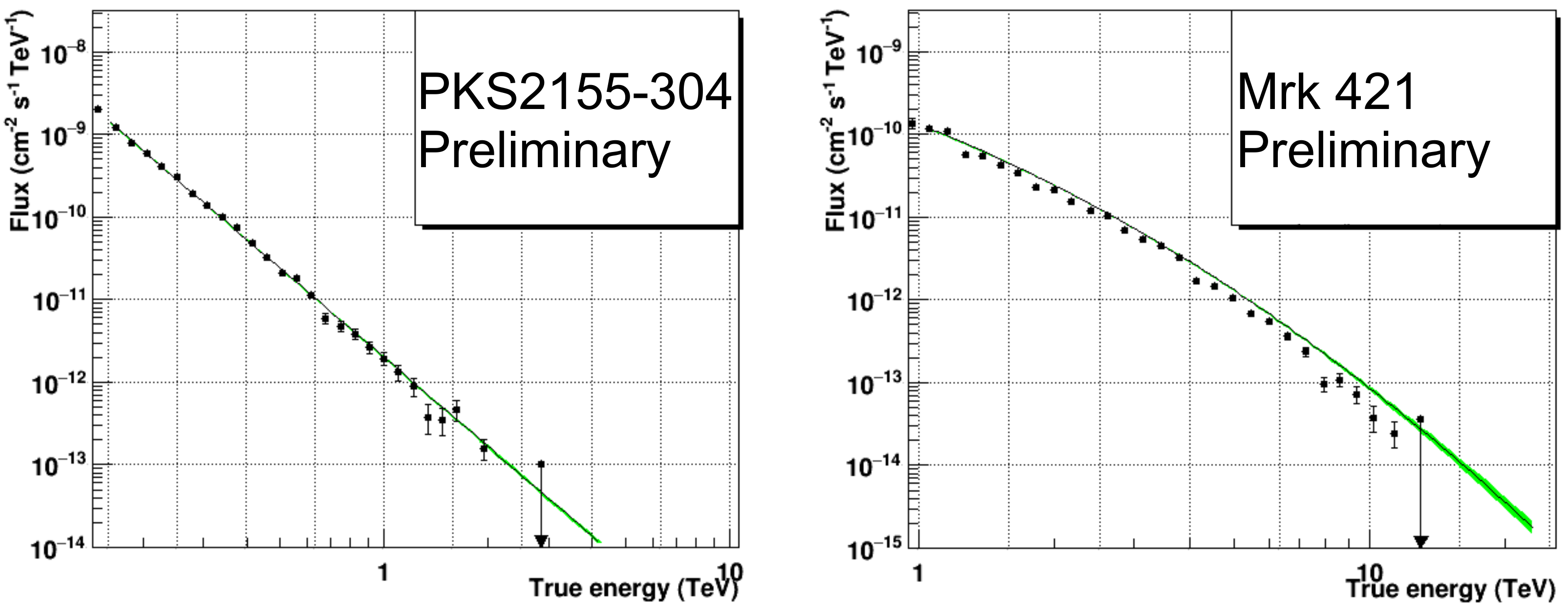}
\caption{Energy Spectra of PKS 2155$-$304 (\textit{left}) and Mrk 421 (\textit{right}) as obtained with IRFs from RWS.}
\label{spectra}
\end{figure}
Spectral fits with the new approach have been carried out for PKS 2155$-$304 and Markarian 421. The results are shown in Figure~\ref{spectra}. As the implementation of the adapted fitting approach still has to be fully validated, the results are to be considered as preliminary.

%
%
%
%

\section{Conclusions}

Here we introduced a novel simulation approach for IACT arrays which accounts for the \textit{run}-specific observation and detector settings. The method has been fully developed and implemented into one of the simulation and analysis frameworks of H.E.S.S. As the concept leads to a more realistic simulation, a reduction of systematic errors is expected. The impact of this expected improvement on the PSF extraction has already been shown in Section~\ref{morphology}. In this regard, morphology measurements with IACTs have been brought to a whole new level. Besides reduced systematic uncertainties, the RWS approach is furthermore computationally more efficient for the simulation of entire IACT data sets with various instrument settings and observation conditions. This will become more and more important for future, increasingly complex IACT arrays, such as the Cherenkov Telescope Array.

%

\bibliographystyle{JHEP}
\bibliography{rws-bib}

\providecommand{\href}[2]{#2}\begingroup\raggedright\begin{thebibliography}{10}

\bibitem{2016_HessUpgrade}
{G..~Giavitto et al.}, {\it {Upgraded cameras for the HESS imaging atmospheric
  Cherenkov telescopes}},  in {\em Society of Photo-Optical Instrumentation
  Engineers (SPIE) Conference Series}, vol.~9908 of {\em SPIE Proceedings},
  p.~99082H, Aug., 2016.

\bibitem{2011_VeritasUpgrade}
A.~{Nepomuk Otte} and {for the VERITAS Collaboration}, {\it {The Upgrade of
  VERITAS with High Efficiency Photomultipliers}},  {\em Proceedings of the
  32nd ICRC, Beijing} (Oct., 2011).

\bibitem{2016_MagicUpgrade}
{J.~Aleksi{\'c} et al.}, {\it {The major upgrade of the MAGIC telescopes, Part
  I: The hardware improvements and the commissioning of the system}},  {\em
  Astroparticle Physics} {\bf 72} (Jan., 2016) 61--75.

\bibitem{2009_Model}
M.~{de Naurois} and L.~{Rolland}, {\it {A high performance likelihood
  reconstruction of {$\gamma$}-rays for imaging atmospheric Cherenkov
  telescopes}},  {\em Astroparticle Physics} {\bf 32} (Dec., 2009) 231--252.

\bibitem{2014_ImPACT}
R.~D. {Parsons} and J.~A. {Hinton}, {\it {A Monte Carlo template based analysis
  for air-Cherenkov arrays}},  {\em Astroparticle Physics} {\bf 56} (Apr.,
  2014) 26--34.

\bibitem{2009_TMVA}
S.~{Ohm}, C.~{van Eldik}, and K.~{Egberts}, {\it {{$\gamma$}/hadron separation
  in very-high-energy {$\gamma$}-ray astronomy using a multivariate analysis
  method}},  {\em Astroparticle Physics} {\bf 31} (June, 2009) 383--391.

\bibitem{2015_VeritasModel}
{S.~Vincent for the VERITAS Collaboration}, {\it {A Monte Carlo template-based
  analysis for very high definition imaging atmospheric Cherenkov telescopes as
  applied to the VERITAS telescope array}},  {\em ArXiv e-prints} (Sept., 2015)
  [\href{http://arxiv.org/abs/1509.0198}{{\tt arXiv:1509.0198}}].

\bibitem{2006_HessCrab}
{F.~{Aharonian} et al.}, {\it {Observations of the Crab nebula with HESS}},
  {\em Astronomy and Astrophysics} {\bf 457} (Oct., 2006) 899--915.

\bibitem{2016_MAGICCrab}
{J.~Aleksi{\'c} et al.}, {\it {The major upgrade of the MAGIC telescopes, Part
  II: A performance study using observations of the Crab Nebula}},  {\em
  Astroparticle Physics} {\bf 72} (Jan., 2016) 76--94.

\bibitem{1994_Kaskade}
M.~P. {Kertzman} and G.~H. {Sembroski}, {\it {Computer simulation methods for
  investigating the detection characteristics of TeV air Cherenkov
  telescopes}},  {\em Nuclear Instruments and Methods in Physics Research A}
  {\bf 343} (Apr., 1994) 629--643.

\bibitem{2014_CT5}
{J.~Bolmont et al.}, {\it {The camera of the fifth H.E.S.S. telescope. Part I:
  System description}},  {\em Nuclear Instruments and Methods in Physics
  Research A} {\bf 761} (Oct., 2014) 46--57,
  [\href{http://arxiv.org/abs/1310.5877}{{\tt arXiv:1310.5877}}].

\bibitem{2015_MAGICCrab}
{J.~Aleksi{\'c} et al.}, {\it {Measurement of the Crab Nebula spectrum over
  three decades in energy with the MAGIC telescopes}},  {\em Journal of High
  Energy Astrophysics} {\bf 5} (Mar., 2015) 30--38.

\bibitem{2014_TC}
{J.~Hahn et al.}, {\it {Impact of aerosols and adverse atmospheric conditions
  on the data quality for spectral analysis of the H.E.S.S. telescopes}},  {\em
  Astroparticle Physics} {\bf 54} (Feb., 2014) 25--32.

\end{thebibliography}\endgroup

\end{document}